\documentstyle[multicol,aps,epsf]{revtex}

\begin{document}

\author{Mario Castro,$^{1,*}$ Angel S\'anchez,$^{2}$ and Francisco 
Dom\'{\i}nguez-Adame$^{1}$}
\address{$^{1}$GISC, Departamento de F\'{\i}sica de Materiales,
Universidad Complutense, E-28040 Madrid, Spain\\
$^{2}$GISC, Departamento de Matem\'aticas,  Universidad 
Carlos III de Madrid, E-28911 Legan\'es, Madrid, Spain}

\title{Lattice model for kinetics and grain size 
distribution in crystallization} 

\maketitle

\begin{abstract}

We propose a simple, versatile and fast  computational model to understand the 
deviations from the well-known Kolmogorov-Johnson-Mehl-Avrami kinetic theory
found  in metal recrystallization and amorphous semiconductor crystallization. 
Our model describes in detail the kinetics of the transformation and  the grain
size distribution of the product material, and is in good agreement with the 
available experimental data.  Other morphological and kinetic features amenable
of experimental observation are outlined, suggesting new directions for further
validation of the model. 

\end{abstract}

\begin{multicols}{2}
\narrowtext

\section{Introduction}

Mechanical, electronic or magnetic properties of many polycrystallline
materials depend not
only on their chemical composition, but also on the kinetic path  of these
materials toward the non-equilibrium state. Recently, the interest on thin film
transistors made of polycrystalline Si and Si--Ge grown by low-pressure
chemical vapor deposition has been driven by the technological development of
active matrix addressed flat-panel displays\cite{Im} and thin film solar
cells.\cite{Bergmann}  With these and similar applications in mind, the
capability to engineer the size and geometry of grains becomes crucial to
design materials with the required properties.

In general, crystallization  of most materials takes place by a nucleation and
growth mechanism:\cite{Zellama} Nucleation  starts with the appearance of
small atom clusters ({\em embryos}). At a certain fixed temperature, embryos
with sizes greater than a critical one become growing nuclei; otherwise, they
shrink and eventually vanish. Such a critical radius arises from the
competition between the surface tension, $\gamma$, and the difference in free
energy between the amorphous and crystalline phases, $\Delta g$, that favors the
increasing of grain volume, yielding an energy barrier that has to be overcome
to build  up a critical nucleus. For a circular grain of radius $r$, 
the free energy takes the simple form
\begin{equation}
\Delta G=2\pi r\gamma-\pi r^2\Delta g.
\end{equation}

The free energy $\Delta G$ has a maximum, the energy barrier, at the critical 
radius $r^{*}=\gamma/\Delta g$. Subsequently, surviving nuclei ($r > r^{*}$)
grow by incorporation of neighboring atoms, yielding a moving boundary with 
temperature dependent velocity that gradually covers the untransformed phase.
Growing grains impinge upon each other, forming a grain boundary, and growth
ceases perpendicularly to that boundary.  Therefore, the structure consists of
vertices connected by edges (grain boundaries) which surround the grains. The
number of edges joined to a given vertex is $3$. In some cases, at high
temperatures these boundaries move until they reach a more favorable
equilibrium configuration (in two dimensions, the equilibrium angles at a
vertex are  $120^{\text o}$).\cite{Atkinson}

In the past few years, the belief that this picture is far too simple to
properly describe nucleation-driven crystallization has progressively spread
among the researchers in the field. This is chiefly due to two problems: On the
one hand, this  theory of nucleation and growth predicts an energy barrier 
much larger than 
the experimental one, implying that nucleation would be hardly probable
at available annealing temperatures.\cite{Doherty}  On the other hand, it is
known that in crystallization of Si over SiO$_2$ substrates, nucleation
develops in the Si/SiO$_2$ interface due to inhomogeneities or impurities that
catalyze the transformation.\cite{Silicon} Therefore, a theory of homogeneous 
nucleation and growth is not entirely applicable to the referred   experiments
as well as to other examples reported in the literature.\cite{Price} 

In addition to the difficulties above, it is clear that   the transformation
kinetics is also problematic. It is generally accepted that the fraction of
transformed material during crystallization, $X(t)$, obeys the 
Kolmogorov-Johnson-Mehl-Avrami (KJMA) model,\cite{KJMA} according to which  
\begin{equation}
X(t)=1-\exp(-at^m),
\label{kjma_eq}
\end{equation}
where $a$ is a nucleation- and growth-rate dependent constant and $m$ is  an
exponent characteristic of the experimental conditions. Two well-defined limits
have been extensively discussed in the literature: When all the nuclei are
present and begin to grow at the beginning of the transformation, the KJMA
exponent, $m$, is equal to $2$ (in quasi-two-dimensional growth like thin
films), and the nucleation  condition is termed {\em site saturation}.  The
product microstructure is tessellated by the so-called Voronoi polygons (or
Wigner-Seitz cells). On the contrary, when new nuclei appear at every step of
the transformation, $m=3$ and the process is named {\em continuous} or {\em
homogeneous nucleation}. Plots of $\log[-\log(1-X)]$ against  $\log(t)$ (called
KJMA plots)  should then be straight lines of slope $m$. Although in some
cases, the KJMA theory explains correctly the transformation kinetics, its
general validity has been questioned in the last few 
years,\cite{Baram,Pineda} and
several papers have been devoted to understand this question in different
ways.\cite{Sessa,Fanfoni,Siclen} However, there are still some open  questions:
An exponent between $2$ and $3$  is experimentally obtained (between $3$ and
$4$ in three dimensions);\cite{Price} the KJMA plots from experimental data do
not fit to a straight line in some cases;\cite{Price} and, finally, the
connection between geometrical properties (grain size distributions) and the
KJMA exponent is not clear.

In this paper we report on a detailed investigation of a probabilistic lattice
model which relates in a clear-cut way  the mentioned problems to the
inhomogeneities in the sample, i.e., the fact that heterogeneous nucleation
takes place. Indeed,  heterogeneous nucleation is rather common in nature due
to impurities or substrate cavities resulting from roughness, among others.
Within our model, the connection  between such heterogeneous nucleation and the
deviations from the simplest nucleation picture become evident. Furthermore, as
we will see below, our model predicts measurable quantities, such as the grain
size distribution or the KJMA exponent, which are in good agreement with the
experiments. Our paper is organized according to the following  scheme: in
Sec.\ II we introduce our model and discuss in depth the relationship between
its defining parameters and physical ones. Section~III collects the results of
an extensive simulation program which establishes the main features of the
model. Finally, Sec.~IV discusses the connection between our model and
experiments, and concludes the paper by summarizing our main findings and
collecting some prospects and open questions.  

\section{The model}

\subsection{Evolution rules}

Our model is based in some previous ideas by Cahn \cite{Cahn1,Cahn2} and
Beck,\cite{Beck} and its key proposal is that the material is not perfectly 
homogeneous but, on the contrary, it contains regions with some extra energy
(regions with some order produced during deposition or substrate impurities) 
at which nucleation is more probable.
Our aim in this section is to provide a
detailed description of our model (largely expanding the preliminary, 
short report presented earlier in Ref.~\onlinecite{Mario_apl}), and how
the basic idea mentioned above is implemented in it.

The model is defined on a two-dimensional lattice (square and triangular
lattices were employed with essentially similar results)  with periodic
boundary conditions; generalizations can straightforwardly be done to any
spatial dimension.  In the beginning ($t=0$),  every lattice site (or node) 
$\mathbf{x}$ belongs to a certain grain or state. We represent the situation at
$\mathbf x$ by $q({\mathbf x},t)=0,1,2,$\ldots, the state $0$ being that of an
untransformed  region. The lattice spacing is therefore the experimental
resolution, usually greater that $r^{*}$. Following the idea that the amorphous
phase has random regions at which nucleation is favored, we choose a fraction
$c$ of the total lattice sites and label those as {\em able to nucleate}. We
term these energetically favorable sites {\em potential} nuclei.  

Simulation proceeds in discrete time steps of duration $\tau$.  The system
evolves by parallel updating according to the following rules and considering
that initially all the material is untransformed, i.e., $q({\mathbf
x},0)=0$ for all lattice sites:

\begin{itemize}

\item An already transformed site remains at the same state forever.

\item An untransformed {\em potential\/} site may become a new non-existing
state ({\em i.e.}, crystallizes) with probability $n$ (nucleation probability)
if and only if there are no transformed nearest neighbors around it.

\item An untransformed site (including potential sites) transforms into an
already existing  transformed state with probability $g$ (growth probability)
if and only if there is at least one transformed site of that type on its
neighborhood. The new state is randomly chosen among the neighboring grain
states, if there are more than one.

\end{itemize}

Note that we have termed $g$ as {\em growth probability} and not {\em growth
rate}.  The actual growth rate is a non-trivial function $f(g)$, because when
$g<1$ the grains grow with a rough boundary. For the model parameters, we
expect a functional form $n\sim e^{-E_{n}/k_BT}$ and $f(g)\sim
e^{-E_{g}/k_BT}$, where $E_n$ and $E_g$ are the energy barriers for nucleation
and for growth, respectively (see below). Hence, temperature is implicit in the
model parameters. We discuss these relationships in depth in the next
subsections.

\subsection{Physically relevant magnitudes}

As we mentioned above, the crystalline fraction is approximately given by 
Eq.~(\ref{kjma_eq}), with some exponent $m$ depending on the dimensionality and
type of nucleation. Experimentally, the crystalline fraction is measured from
the intensity of the peaks of X-ray diffraction of the microstructure as
material transforms from  the amorphous to the polycrystalline phase. In the
following, we will assume that there is not any preferential
direction,\cite{Comentario_Si} that is, $n$ is the same for all potential
sites, and $g$ is the same for all grains.

The other experimentally measurable magnitude is the grain size distribution,
$P(A)$, defined as the fraction of grains with a given area $A$. To compare
with simulation results, we will usually plot the normalized distribution of
reduced area $A^\prime=A/\bar{A}$, where $\bar{A}$ is the mean area:
\begin{equation}
\bar{A}=\int_0^\infty AP(A)dA.
\end{equation}
This distribution changes dramatically with nucleation
conditions.\cite{Thompson}
Some of the available experimental data are given in terms of the distribution
of grain diameters, $P(d)$. As we will demonstrate below, this distribution is
equivalent to the distribution of {\em effective diameter} $A^{1/2}/\pi$ (or
simply $A^{1/2}$), which is computationally less expensive to calculate. 
Hence, we will present our results in terms of the effective diameter.

\subsection{Time and length scales}
\label{timescales}

To begin with, let us show that the potential sites, distributed randomly
throughout the system, define a characteristic length given by the probability
distribution of nearest  neighbors. Suppose we have $N$ randomly potential
sites in a $L\times L$ system. The mean concentration of potential sites is
$c=N/L^2$. We may ask about the probability of finding a number $k\leq N$ of
these sites in a region of area $A$. This probability is given by the  binomial
distribution:
\begin{equation}
P_k(N,A)=\bigg(\begin{array}{ll}N\\k\\\end{array}\bigg)p^k(1-p)^{N-k},
\label{4}
\end{equation}
where $p=A/L^2$.

Taking the limit $L\to \infty$, $N\to \infty$, while keeping $N/L^2\equiv c$, 
the expression (\ref{4}) tends to the Poisson probability distribution,
\begin{equation}
P_k(A)=\frac{(Ac)^k}{k!}e^{-Ac}.
\end{equation}
If, as stated above, we suppose that the system is isotropic, we may write
the probability of finding $k$ potential sites in a circle of radius $r$ as:
\begin{equation}
P_k(r)=\frac{(\pi r^2c)^k}{k!}e^{-\pi r^2c}.
\label{poisson}
\end{equation}
So, if Eq.\ (\ref{poisson}) is the probability of finding $n$ potential sites
in a disc of radius $r$, then the probability of finding no grains is
\begin{equation}
P_0(r)=e^{-\pi r^2c},
\label{poisson0}
\end{equation}
and the probability of finding at least one neighbor at a distance less than
$r$ is $1-P_0(r)$. This is precisely the probability distribution of nearest
neighbors. In other words, we can obtain the probability density of  
finding at least one neighbor between $r$ and $r+dr$ as follows
\begin{equation}
p(r)dr={d\phantom{r}\over dr}\Big(1-P_0(r)\Big)\,dr=2\pi rce^{-\pi cr^2}dr.
\label{nn_distr}
\end{equation}
The first moment of the distribution is the mean distance among  potential
sites
\begin{equation}
d_m=\int_0^\infty rp(r)dr=c^{-1/2}.
\label{dm}
\end{equation}

On the other hand, the grains grow with constant velocity. For definiteness,
let us take the growth probability $g$ to be  $1$; we will see below that the
results of simulations for other values of $g$ can be reproduced from 
simulations with $g=1$ conveniently rescaled. With this choice, the grain
radius grows according to the law  $r(t)=  \Omega t$,  where $\Omega$ is a
geometrical coefficient that depends on the underlying lattice. Thus, we may
define the mean time at which the growing grains will impinge, or {\em
overlap} time, as $\Omega t_o=c^{-1/2}$ or, in general, i.e., 
ignoring the details of the lattice, $t_o\sim c^{-1/2}$.

The characteristic time scale arising from the concentration of nucleation
sites is not the only one:
Indeed, the nucleation probability defines another characteristic time.
Being more specific, the number of sites that have nucleated per unit time is
proportional to the available ones
\begin{displaymath}
\frac{dN(t)}{dt}=n[N_{\text{max}}-N(t)],
\end{displaymath}
where $N_{\text{max}}=cL^2$. Thus, we have
\begin{equation}
N(t)=N_{\text{max}}(1-e^{-nt}) \Rightarrow \rho(t)=c(1-e^{-nt}),
\end{equation}
$\rho(t)$ being the concentration of already nucleated potential sites at time
$t$. In view of this, we define the characteristic nucleation time $t_n=1/n$. 
As we will see below, the competition between time scales characterizes the
final microstructure.

In a general case, some of the potential sites will be covered by other growing
grains and therefore their nucleation is inhibited. The mean distance of the
potential sites that become actual grains is, replacing $c$ by $\rho(t)$ in
Eq.~(\ref{nn_distr}),
\begin{equation}
d_m(t)=\frac{1}{\sqrt{\rho(t)}}=\frac{1}{c^{1/2}(1-e^{-nt})}.
\end{equation}

If $t_n\ll t_o$, almost every potential site nucleates before grains impinge
upon each other. We term this situation {\em fast nucleation\/}, and in terms
of our model parameters it means that $n\gg c^{1/2}$. This situation is similar
to site saturation nucleation, in which every potential site nucleates at
$t=0$. The KJMA exponent will be close to $2$ and the grain size distribution
will be similar to that of site saturation. Note that, when $n=1$, the exact
limit is obtained for every  concentration $c<1$, but concentrations $c$ close
to $1$ yield a mean grain size of just a few  times the critical radius, $r^*$,
which in fact has not much to do with the experimentally measured values. In
this case, $t_n$ is approximately equal to the simulation time step, $\tau$, 
so the characteristic time scale is $\tau_{\text{fast}}\sim t_o\sim c^{-1/2}$.

Analogously, if $t_n\gg t_o$ then $c^{1/2}\gg n$ and growing grains will
overlap potential sites before these have nucleated, forcing the number of
nucleating grains to decrease with time. As new grains still appear at every
stage of the transformation, we expect approximately  homogeneous nucleation,
and correspondingly a  KJMA exponent close  to $3$. We term this situation {\em
slow nucleation}. Comparing the radii of the grains with the mean distance
among them we find the characteristic time of the process:
\begin{eqnarray*}
\frac{1}{c^{1/2}[1-\exp(-n\tau_{\text{slow}})]}& \simeq &
\frac{1}{(cn\tau_{\text{slow}})^{1/2}}\sim \\ &\sim & r(\tau_{\text{slow}})=\Omega
\tau_{\text{slow}},
\end{eqnarray*}
and hence
\begin{equation}
\tau_{\text{slow}}\sim \frac{1}{(cn)^{1/3}}.
\end{equation}
The important point, however, is the fact that between both limits we will find
a wide range of KJMA exponents and grain size distributions, consistently with
the experimental results.

\section{Numerical results and discussion}

\subsection{Isolated grain shapes}

\begin{figure}[!htp]
\epsfxsize=8cm
\epsffile{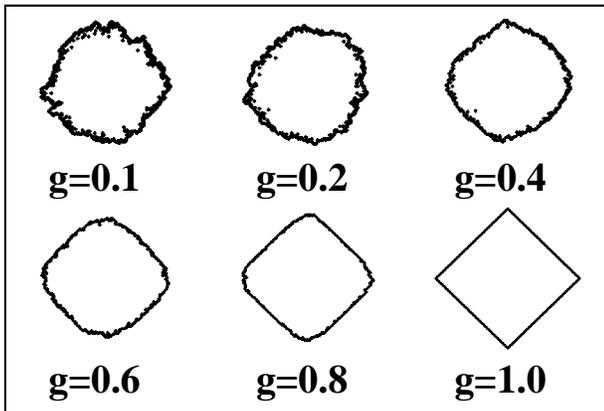}
\caption{Individual grains grown on a square lattice for different growth 
probabilities $g$.}
\label{square_g}
\end{figure}

\begin{figure}[!htp]
\epsfxsize=8cm
\epsffile{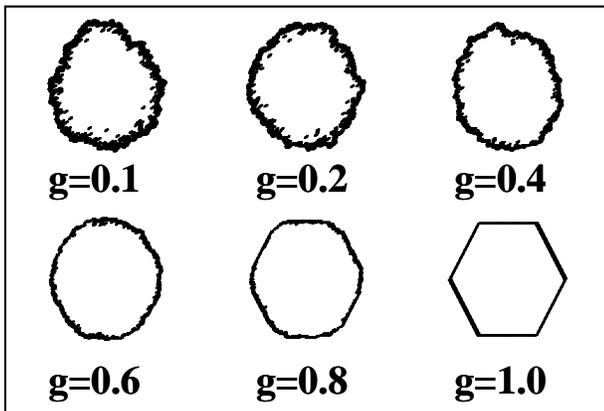}
\caption{Individual grains grown on a triangular lattice for different growth 
probabilities $g$.}
\label{triangular_g}
\end{figure}

An isolated grain, i.e., a grain  completely surrounded by untransformed
material, grows isotropically. Thus, in a continuum medium, the grain boundary
is nearly a circumference.  Nevertheless, the shape of such propagating
interfaces in our model depends strongly on the underlying lattice. For
example, in the limit case $g=1$, a grain growing in a square lattice is 
square shaped, whereas if growing in a triangular lattice it is hexagonal
shaped. As the growth probability $g$  diminishes, the underlying lattice
effects seem to vanish, and grains are approximately 
circular, with a rough boundary.  In
Figs.~\ref{square_g} and~\ref{triangular_g} we show the dependence of the grain
shape on the growth probability, varying $g$ from $0.1$ to $1$, on square and
triangular lattices respectively. We see that for $g\lesssim 0.4$ the shape 
of an isolated growing grain becomes practically independent of the lattice, 
whereas for larger values of $g$, the grain shape exhibits the influence of 
the lattice geometry. 
It is important to note that this does not occur when many grains grow 
simultaneously, as in this case the grain geometry is determined by the 
succesive impingement with its neighbors.  

\begin{figure}[!htp]
\epsfxsize=8cm
\epsffile{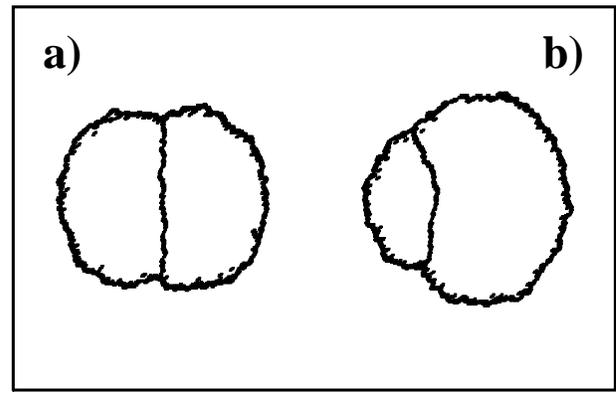}
\caption{Boundaries between two individual grains
obtained from simulations with $g=0.5$ on a triangular lattice: (a) both 
grains nucleate at the same time, and (b) they nucleate at different times,
yielding a curved interface.
}
\label{recta_hiperbola}
\end{figure}

In connection with the last remark, it is interesting to consider
another issue related to boundaries, namely that of boundaries between
different grains. Let $r_1$ and $r_2$ be the radii of two circular grains;
their boundary is then defined by the equation\cite{Thompson}
\begin{equation}
r_1+v_gt_1=r_2+v_gt_2,
\end{equation}
where $v_g$ is the growth velocity and $t_{1,2}$ the elapsed time since each
grains nucleated. When grains started to grow at the same time ($t_1=t_2$), the
boundary is a straight line. Otherwise, it is a hyperbola. In
Fig.~\ref{recta_hiperbola} we plot two examples
of interfaces in which, in spite of
the fact that interfaces are noisy, both characteristic curves are revealed.

\subsection{Kinetics}

\begin{figure}[!htp]
\epsfxsize=8cm
\epsffile{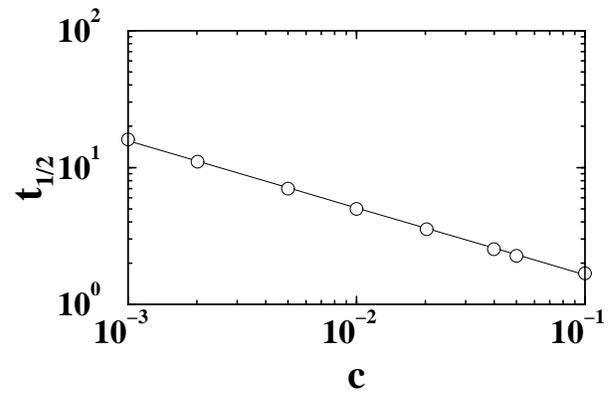}
\caption{$\log-\log$ plot of the characteristic 
time $t_{1/2}$ {\em vs.} $c$ in 
the fast nucleation limit over a square lattice: ($\circ$) Simulation; solid
line:   power-law fit with  slope  $-0.50\pm 0.01$.}
\label{tau_fast}
\end{figure}

\begin{figure}[!htp]
\epsfxsize=8cm
\epsffile{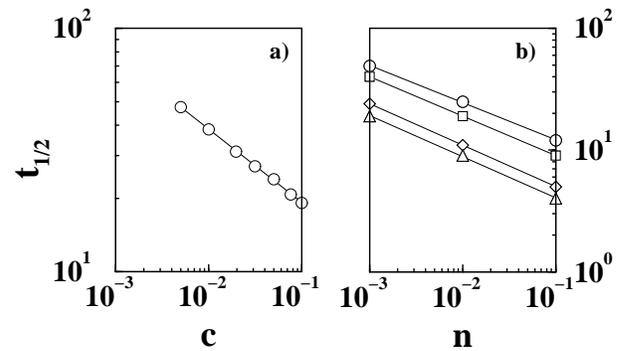}
\caption{$\log-\log$ plot of the characteristic time $t_{1/2}$: (a) ($\circ$)
Simulation value, solid line is a power-law fits with slope $-0.32\pm 0.01$;
(b) Symbols stand for simulation, solid lines are power fittings:
($\circ$) $c=0.005$, slope: $0.34\pm 0.02$;($\Box$) $c=0.01$, slope: 
$0.34\pm 0.01$; ($\Diamond$) $c=0.05$, slope: $0.32\pm0.01$; and ($\triangle$)
$c=0.1$, slope: $0.31\pm0.02$.}
\label{tau_slow}
\end{figure}

We have simulated $1000\times1000$ triangular and square lattices and averaged 
the outcome of $50$ different realizations for each choice of parameters 
(characteristic simulation times are about $15$ to $45$ minutes in a Pentium II
personal computer).  The crystalline fraction ranges from $0$ to $1$, so  we
define the typical simulation (or experimental) time as the time $t_{1/2}$ at
which  $X(t_{1/2})=1/2$. As a check on our ideas, we begun by 
verifying the dependence of this parameter on the time scales defined 
above. In Figs.~\ref{tau_fast} and~\ref{tau_slow}, we plot
$t_{1/2}$ for different parameters  in the fast and slow nucleation limits. A
very good agreement is observed with the expected behavior of $t_{1/2}\sim
\tau_{\text{fast}}$ and $t_{1/2}\sim \tau_{\text{slow}}$  discussed in
Sec.~\ref{timescales}. Therefore, we can be confident that the expectations 
drawn above about the behavior of the model, based on theoretical 
considerations, will be fulfilled. 

\begin{figure}[!htp]
\epsfxsize=8cm
\epsffile{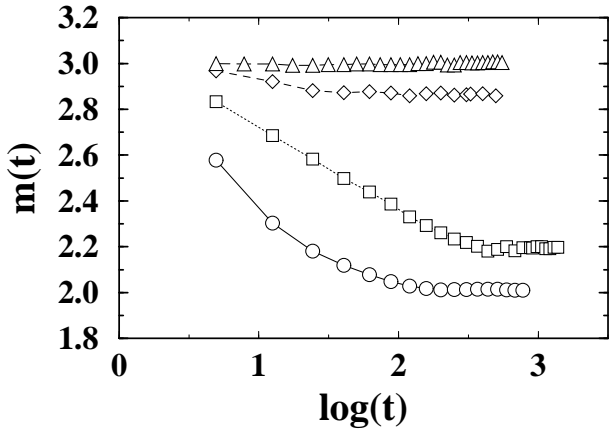}
\caption{Transient KJMA exponent {\em vs.} $\log(t)$. Circles: $n=1$
and $c=0.001$; squares: $n=0.5$ and $c=0.005$; diamonds: $n=0.1$
and $c=0.05$ and triangles: $n=0.01$ and $c=0.1$. $g=1$ in all cases.}
\label{derivada_kjma}
\end{figure}

\begin{figure}[!htp]
\epsfxsize=8cm
\epsffile{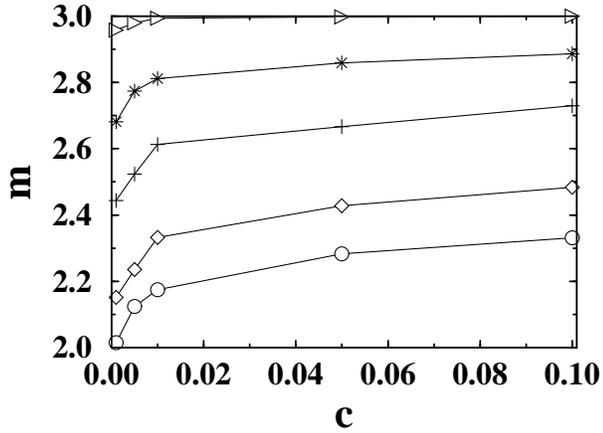}
\caption{KJMA exponent dependence on the concentration probability $c$ on
a $1000\times 1000$ triangular lattice. From top to bottom: $n=0.001$,
$n=0.01$, $n=0.03$, $n=0.07$ and $n=1$.}
\label{kjma_vs_c_tri}
\end{figure}

\begin{figure}[!htp]
\epsfxsize=8cm
\epsffile{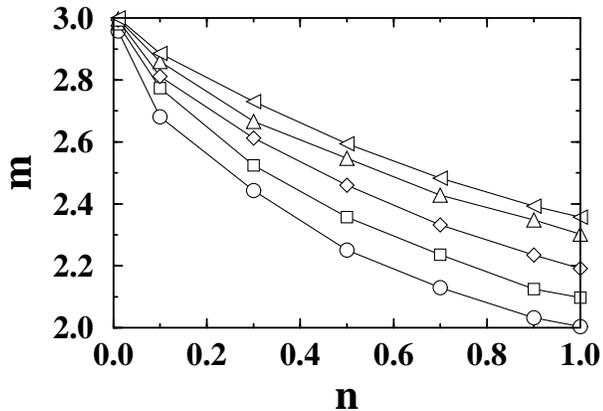}
\caption{KJMA exponent dependence on the nucleation probability $n$ on
a $1000\times 1000$ triangular lattice: ($\circ$) $c=0.001$;
($\Box$) $c=0.005$; ($\Diamond$) $c=0.01$; ($\triangle$) $c=0.05$; and 
($\lhd$) $c=0.1$, }
\label{kjma_vs_n_tri}
\end{figure}

The first key feature to analyze relates to the crystallization kinetics 
as seen through 
KJMA plots. Our results show that
those are not the straight lines predicted by the KJMA model: This can 
be best seen by looking at the 
transient KJMA exponent, defined as
\begin{equation}
m(t)=\frac{d}{d(\log t)}\left[\log(-\log(1-X))\right],
\end{equation}
{}Figure~\ref{derivada_kjma} shows that the KJMA exponent always 
decreases from its initial value to an asymptotic, time independent one;  
correspondingly and in agreement with the experiments, KJMA plots 
approach straight lines only at late times.
We note that, in determining $m(t)$, care has to be taken 
from the computational point of view as in some cases the number of
steps needed to complete the transformation is too short.  In addition, 
it is necessary to remove the last few instants of the time evolution,
as they exhibit large
finite size effects. The asymptotic value is the one we take
from simulations and the one
plotted in Fig.~\ref{kjma_vs_c_tri} showing the dependence of the KJMA
exponent with the potential site concentration $c$. Alternatively, Fig.\
\ref{kjma_vs_n_tri}  depicts the dependence of KJMA exponent on the nucleation
probability $n$. We thus see that there is a large variability of the 
KJMA exponent, covering all the range between 2 and 3 in this two-dimensional
case, that depends on the relationship between the nucleation probability $n$
(i.e., the nucleation rate) and the concentration of nucleation sites $c$. 
This result is a step beyond KJMA theory, and agrees 
with the fact that experiments offer very 
different results, with exponents between 2 and 3. 

\subsection{Grain area and grain diameter}
\label{areas_sec}

\begin{figure}[!htp]
\epsfxsize=8cm
\epsffile{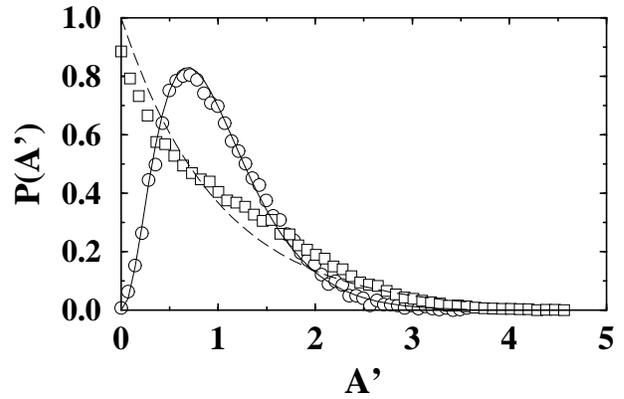}
\caption{Grain reduced area distribution: ($\circ$) Simulation with $n=1$,
$g=1$ and $c=0.001$; ($\Box$) simulation with $n=0.001$, $g=1$ and $c=0.5$.
Solid line: exact value from Eq.\ (\ref{Weire_eq}); dashed line: from
Eq.\ (\ref{Mulheran_eq}).}
\label{fitting}
\end{figure}

In order to further check the model results, we have to compare the grain size
distributions with some well-accepted theoretical ones. Although these 
distributions are obtained phenomenologically, the agreement with experiments
and simulations  is very good. Under some assumptions about the mean
number of neighbors of a nucleation center, Weire {\em et al.}\ proposed a
simple distribution for site saturation \cite{Weire}
\begin{equation}
P(A^\prime)=(A^\prime)^{\alpha-1}\alpha^\alpha \exp[-\alpha
A^\prime]/\Gamma(\alpha),
\label{Weire_eq}
\end{equation}
where $\alpha \simeq 3.65$, and $A^\prime=A/\bar{A}$ is the reduced area. In
Fig.~\ref{fitting} we plot the normalized grain size distribution (circles) 
for different parameters for which $m\simeq 2$, i.e., site saturation, and
compare it with Eq.~(\ref{Weire_eq}) (solid line).

Similarly, in the case of homogeneous nucleation, a simple (but not so
accurate) expression has been proposed\cite{Mulheran}
\begin{equation}
P(A^\prime)= \exp[-A^\prime].
\label{Mulheran_eq}
\end{equation}
Our model shows some slight deviations from this equation, as seen in 
Fig.~\ref{fitting}. Interestingly, these are the same as in other model
simulations,\cite{Mulheran} and, in addition, we have to keep in mind the
applicability limitations of Eq.~(\ref{Mulheran_eq}).\cite{Mulheran}
Therefore, we believe that the behavior displayed by our model is also 
fully satisfactory in this limit.

\begin{figure}[!htp]
\epsfxsize=8cm
\epsffile{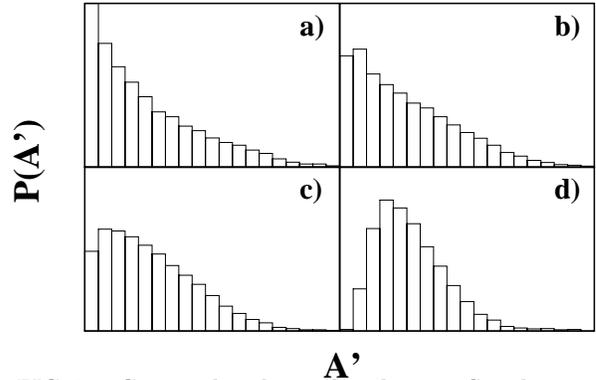}
\caption{Grain reduced area distribution. Simulation with: a) $n=0.01$
and $c=0.1$; b) $n=0.1$ and $c=0.05$; c) $n=0.5$
and $c=0.005$  and d) $n=1$ and $c=0.001$. 
$g=1$ in all cases. Horizontal axis ranges from
$0$ to $4$ and vertical axes from $0$ to $1$ in four graphs.}
\label{Evol}
\end{figure}

Once we have checked the validity of the model in the well-known limits, we
report on the influence of the 
nucleation probability, $n$, and the potential site
concentration, $c$, on the grain size distribution. In Fig.~\ref{Evol} we plot
several grain size distributions when we pick both parameters along a line
going from the slow to the fast nucleation limit. In so doing, we cross from a
extended distribution to a stretched one, as we would expect in view of 
Eqs.\ (\ref{Weire_eq}) and (\ref{Mulheran_eq}).

\begin{figure}[!htp]
\epsfxsize=8cm
\epsffile{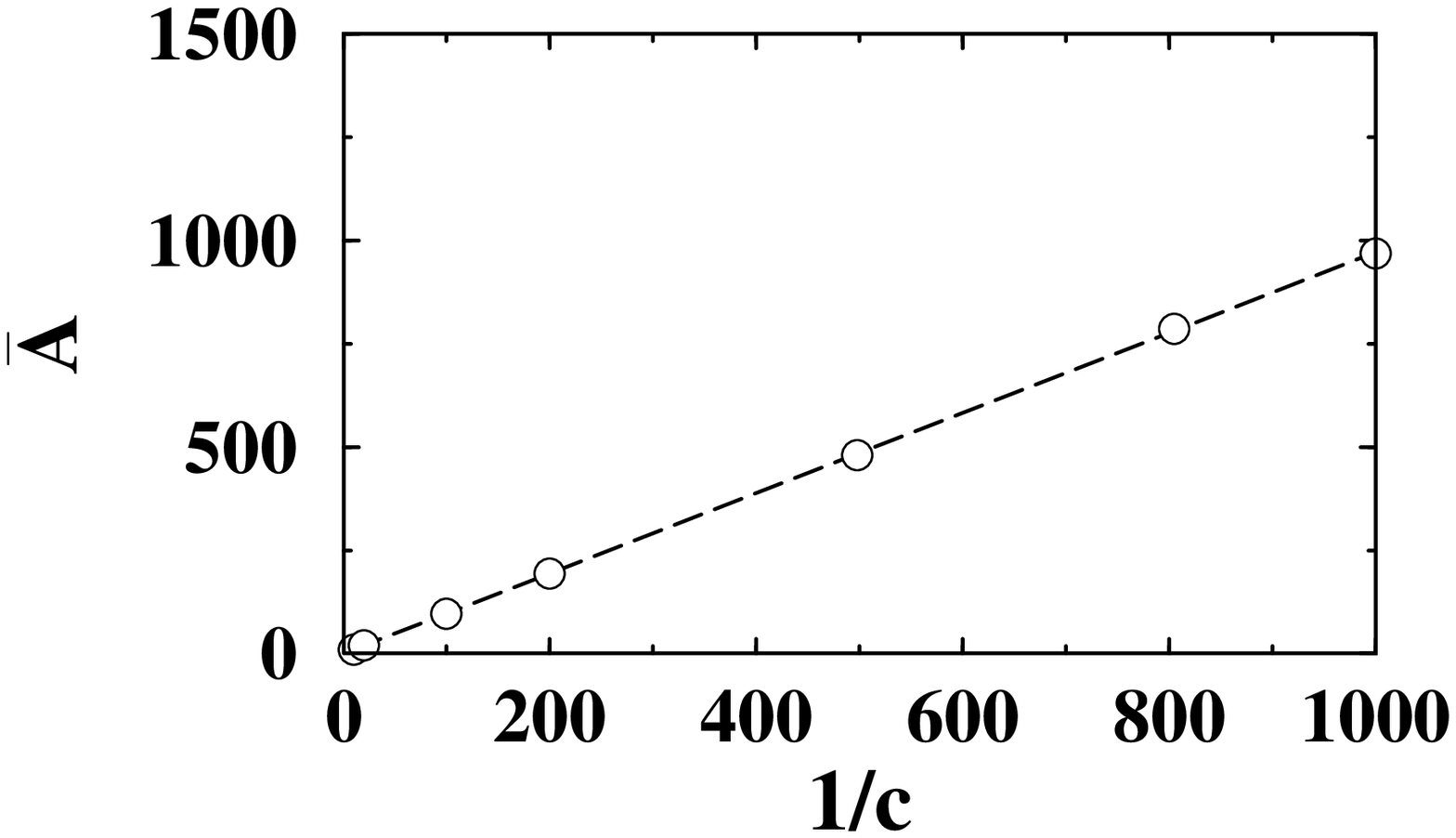}
\caption{Mean area {\em vs} inverse of the potential site concentration.
($\circ$) Simulation values. Dashed line is a linear fit.}
\label{mean_diam_c}
\end{figure}

\begin{figure}[!htp]
\epsfxsize=8cm
\epsffile{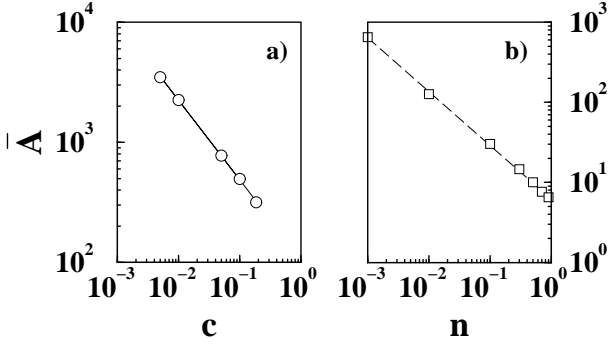}
\caption{$\log-\log$ plot of the mean area $\bar{A}$: (a) ($\circ$)
Simulation value; solid line: power-law fit with slope $-0.66\pm 0.01$;
(b) ($\Box$) simulation, dashed line: power-law fit with slope 
$-0.67\pm 0.02$.}
\label{mean_diam_n}
\end{figure}

Let us now turn to the issue of the mean grain size. 
As we have pointed out, in the fast nucleation limit the characteristic length
scale is related to the mean potential site distance, $c^{-1/2}$:  In this
case, we expect the mean grain diameter to be proportional to that scale. In
Fig.~\ref{mean_diam_c} we show this linear dependence of the mean area on
$c^{-1}$.  On the contrary, in the slow nucleation limit, when the
concentration $c$ is relatively large, the  grains grow on an {\em effective}
homogeneous medium. Roughly speaking, the mean distance among potential sites
is so small that the grain radii is very soon larger than this distance. Thus,
the  characteristic length scale, $d$, is that of the grains when they impinge
upon each other. As the grain radius grows linearly with time, we expect 
$d\sim t_{1/2}$, so $\bar{A}^{1/2}\sim (cn)^{-1/3}$ and $\bar{A}\sim
(cn)^{-2/3}$. Figure \ref{mean_diam_n} confirms that this 
simple analysis is very accurate.

\begin{figure}[!htp]
\epsfxsize=8cm
\epsffile{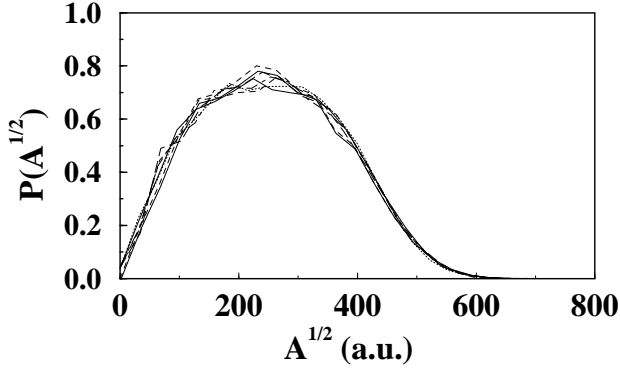}
\caption{Collapse of the 
grain effective diameter normalized distributions for eight 
$g$ values ranging from $0.05$ to $1$.}
\label{colapso_g}
\end{figure}

Finally, there are two questions we announced in Sec.\ II whose 
validation has been left postponed. We now address these points, 
beginning by that of the effect of the parameter $g$, which so
far we have restricted to $g=1$. For every value of $g$,
the growth rate, $f(g)$, defines a characteristic time related to
the temporal scale at which the grains spread on the amorphous substrate.   
Thus, we expect that by
rescaling the simulation time step $\tau \rightarrow f(g)\tau$ (with
$f(g)\rightarrow \Omega$, as $g \rightarrow 1$, $\Omega$ being the 
geometrical coefficient introduced in Sec.\ IIC), the mean grain
size will depend only on the ratio $n/f(g)$. We have not 
been able to  obtain an
analytical expression for $f(g)$ but we can  calculate it numerically for the
required $g$, by growing an isolated grain. In Fig.~\ref{colapso_g} we show
the excellent collapse of different effective  diameter distributions for
several couples $(n,g)$ with constant $n/f(g)$. This result shows that 
the outcome of the simulations reported here for $g=1$ truly represents, 
except for a factor, the model characteristics for other values of $g$. 

\begin{figure}[!htp]
\epsfxsize=8cm
\epsffile{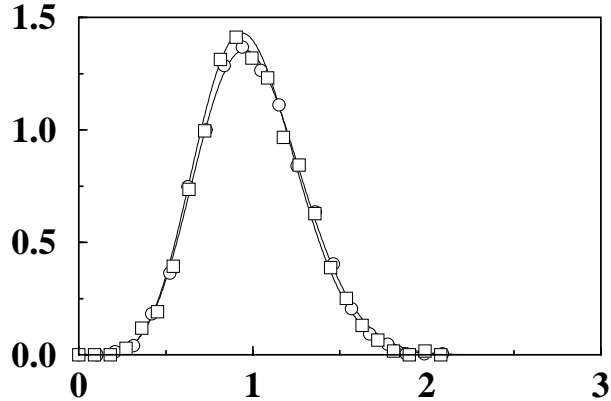}
\caption{Numerical comparison between normalized distributions of
reduced grain diameter ($\circ$), $d^\prime$, and  reduced effective 
diameter, $(A^\prime)^{1/2}$ ($\Box$).}
\label{effective_vs_real}
\end{figure}

The other pending question is related to the mean grain diameter. 
So far, we have discussed our results in terms of the mean grain area or the
mean effective diameter size.  To verify whether the
effective diameter distribution is the same as the real diameter distribution,
which is computationally much more demanding, 
we have compared them in several cases. The comparison is shown in 
Fig.\ \ref{effective_vs_real}, by plotting
the referred normalized distributions. The
correlation between both sets of points is greater than $99.9\%$, 
allowing us to conclude that the reports above in terms 
of areas carries over to the mean diameter picture without significant
changes. 

\subsection{Mean number of neighbors}

\begin{figure}[!htp]
\epsfxsize=8cm
\epsffile{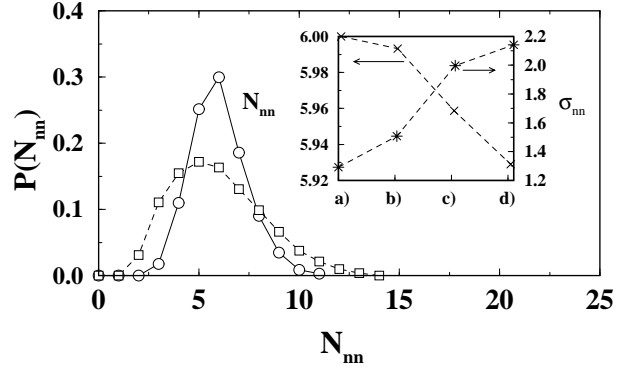}
\caption{Nearest neighbor number normalized distribution. ($\circ$) $n=1$,
$c=0.001$ (site saturation); ($\Box$) $n=0.01$ and $c=0.1$ (homogeneous
nucleation). Inset: Mean number of neighbors, $N_{nn}$, and its variance,
$\sigma_{nn}$ with: a) $n=0.01$ and $c=0.1$; b) $n=0.1$ and $c=0.05$; c) $n=0.5$
and $c=0.005$  and d) $n=1$ and $c=0.001$.} 
\label{vecinos}
\end{figure}

Some theoretical approaches to equilibrium crystallized configurations deal
with the mean number of neighbors, $N_{nn}$,\cite{Mulheran} or equivalently,
considering the final product as a polygon tessellation of space,  the mean
number of sides of those polygons. If the material is divided in equal size
hexagons, this distribution is  $P(N_{nn})=\delta(N_{nn}-6)$. In 
Fig.~\ref{vecinos} we plot the  numerical distribution of nearest neighbors for
site saturation and homogeneous nucleation. The asymmetry and the variance of
the mean number of neighbors are the main differences in both limits. The inset
in Fig.~\ref{vecinos} shows  the mean number of neighbors and the
corresponding  changes in variance for different parameters. Clearly, the
distribution spreads out and loses its simmetry in homogeneous nucleation. 
Furthermore, computing the mean  number of neighbors against the nucleation
time for all of the grains in the sample we find that the {\em younger} grains
have less number of sides than the {\em older\/} ones, which explains this
asymmetry.  Hence, this distribution can be another element of comparison with
experiments. We remark that secondary crystallization (or abnormal grain
growth) is due to these deviations from the ideal configuration.

\subsection{Temperature and applicability of the model}
\label{sec_temp}

To conclude our analysis of heterogenous nucleation, we present some results of
the influence of temperature in product properties. In addition, this will 
allow us to show that the model gives consistent results when realistic
parameters are chosen to reproduce an actual material. 
The mean grain area in
homogeneous nucleation of two-dimensional disks is given by the simple
relation:\cite{Thompson}
\begin{equation}
\bar{A}\sim\left(\frac{G_0}{N_0}\right)^{2/3},
\label{dostercios}
\end{equation}
where $N_0$ and $G_0$ are the nucleation and growth rates  respectively.
Identifying $N_0$ with $n$ and $G_0$ with $f(g)$, we have obtained similar 
results (see Sec.~\ref{areas_sec}). As nucleation and growth are activated
processes, we postulate an Arrhenius-like dependence of nucleation and growth
probabilities:
\begin{equation}
n \sim \exp[-E_n/k_BT ],\quad f(g) \sim \exp[-E_g/k_BT].\nonumber
\end{equation}
In homogeneous nucleation, as we have reported, we can redefine $n$ and $g$ to
set $g=1$; hence, the temperature is introduced in our model by means of the
nucleation probability
\begin{equation}
n\rightarrow n^\prime=n/f(g)=n^\prime_0 \exp[-(E_n-E_g)/k_BT],
\label{activa}
\end{equation}
and $g=1$.

\begin{figure}[!htp]
\epsfxsize=8cm
\epsffile{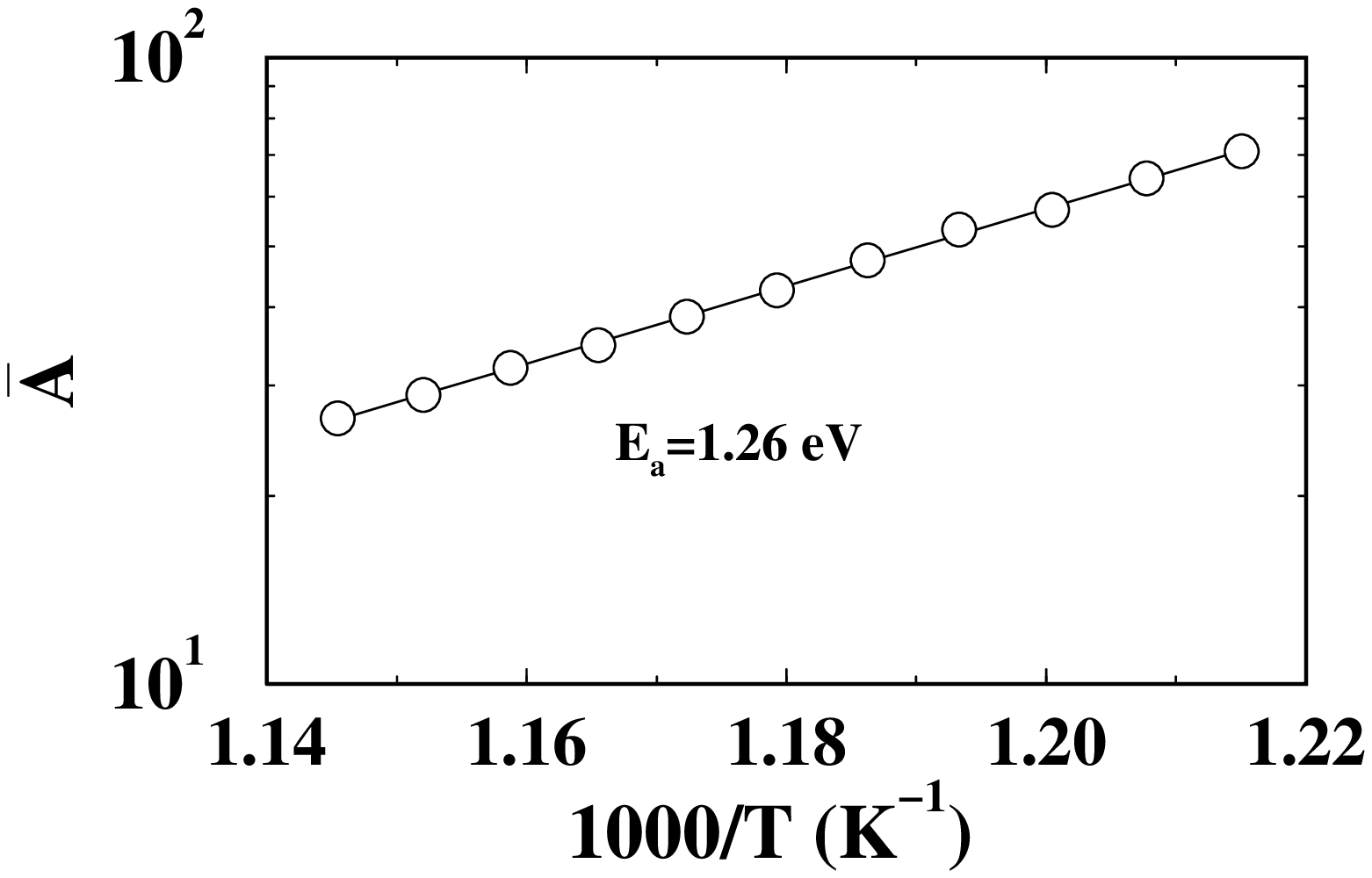}
\caption{Selfconsistency of Eq. (\ref{activa})
with $c=1$ (homogeneous nucleation). 
($\circ$) Simulation; solid line:
exponential fit which gives an activation energy 
$E_a=2(E_n-E_g)/3=1.26\pm0.01$, consistent with $E_n=5.1$ eV and $E_g=3.2$ eV.}
\label{test_temperatura}
\end{figure}

As an example,
if we want to model nondendritic Si crystallization, we may use the
experimental activation energies \cite{Si_energy}: $E_n=5.1$ eV and $E_g=3.2$
eV. Then, $\bar{A}\sim \exp[E_a/k_BT]$, where from  Eq.~(\ref{dostercios})
$E_a=2(E_n-E_g)/3\simeq 1.27$ eV. In Fig.~\ref{test_temperatura} we plot the
mean grain size {\em vs.} $1000/T$. The slope gives $E_a=1.26\pm0.01$ eV, which
is consistent with the introduced values. Thus, the model provides
a simple tool to analyze crystallization experiments: Setting the
activation energies as the program input, we just have to choose a realistic
value of $n^\prime_0$ ({\em e.g.}, in terms of the final number of grains) and
tune the degree of heterogeneities, $c$, in order to compare with 
the experiments. 

\section{Conclusions}

As we have seen, the model proposed in this paper 
provides very accurate and detailed spatial and temporal information
about the system evolution: Crystalline fraction, mean grain area, KJMA
exponent, or mean number of neighbors. The main features observed in 
experiments, such as non-integer KJMA exponents or different types of
grain size distributions are very well reproduced by the model. 
We must conclude, then, that the model captures all the physical 
ingredients involved in the crystallization process: In particular, 
it points out to the inhomogeneity of the nucleation phenomenon (which 
can arise because of the structure of the amorphous material itself, 
or because of defects at the substrate-material interface, for instance)
as the key feature governing the crystallization kinetics and the resulting
grain textures. In view of this, we propose this
model, very unexpensive in terms of computing time, 
as a versatile way to incorporate other physical ingredients as boundary
migration, preferential grain growth or diffusion-controlled growth
which will be the aim of further work.
Finally, from
the experimental perspective, it has to be mentioned 
that the model should be able to explain and
predict some results.  Predictions can be made by means of
$n^\prime$, controlled by changing the annealing  temperature (see
Sec.~\ref{sec_temp}), and $c$ by ion implantation of nucleation centers, or by
some induced impurities or defects  on the sample substrate. Some ordered
distributions of defects can be induced by ion implantation with an
appropriate mask, which can be trivially introduced in our model. These
ideas call for further experimental work in order to confirm the 
validity of our model. 

\acknowledgements
This work was supported by CAM (Madrid, Spain)
under project 07N/0034/98 and by DGESIC (Spain) under
project PB96-0119.

\end{multicols}

\begin{references}
\bibitem{Comentario_Si}	In polycristalline silicon, it  has been observed that
	direction $(111)$ is prefered throughout the transformation, and that
	it mainly nucleates at the interface with the underlying substrate.
\end{references}
\end{document}